# A complete formalized knowledge representation model for advanced digital forensics timeline analysis

**Yoan Chabot a, b, * , Aurelie Bertaux a , Christophe Nicolle a , M-Tahar Kechadi b**

a CheckSem Team, Laboratoire Le2i, UMR CNRS 6306, Faculte des sciences Mirande, Universite de Bourgogne, BP47870, 21078 Dijon, France

b School of Computer Science & Informatics, University College Dublin, Belfield, Dublin 4, Ireland

## Abstract

Having a clear view of events that occurred over time is a difficult objective to achieve in digital investigations (DI). Event reconstruction, which allows investigators to understand the timeline of a crime, is one of the most important step of a DI process. This complex task requires exploration of a large amount of events due to the pervasiveness of new technologies nowadays. Any evidence produced at the end of the investigative process must also meet the requirements of the courts, such as reproducibility, verifiability, validation, etc. For this purpose, we propose a new methodology, supported by theoretical concepts, that can assist investigators through the whole process including the construction and the interpretation of the events describing the case. The proposed approach is based on a model which integrates knowledge of experts from the fields of digital forensics and software development to allow a semantically rich representation of events related to the incident. The main purpose of this model is to allow the analysis of these events in an automatic and efficient way. This paper describes the approach and then focuses on the main conceptual and formal aspects: a formal incident modelization and operators for timeline reconstruction and analysis.

## Introduction

Due to the rapid evolution of digital technologies and their pervasiveness in everyday life, the digital forensics field is facing challenges that were anecdotal a few years ago. Existing digital forensics toolkits, such as EnCase or FTK, simplify and facilitate the work during an investigation. However, the scope of these tools is limited to collection and examination of evidence (i.e. studying its properties), which are the first two steps of the investigation process, as defined in Palmer (2001). To extract acceptable evidence, it is also necessary to deduce new knowledge such as the causes of the current state of the evidence (Carrier and Spafford, 2004b). The field of event reconstruction aims at solving this issue: event reconstruction can be seen as a process of taking as input a set of events and outputting a timeline of the events describing the case. Several approaches have been proposed to carry out event reconstruction, which try to extract events and then represent them in a single timeline (super-timeline) (Gudhjonsson, 2010)). This timeline allows to have a global overview of the events occurring before, during and after a given incident. However, due to the number of events which can be very large, the produced timeline may be quite complicated to analyse. This makes the interpretation of the timeline and therefore the decision making very difficult. In addition, event reconstruction is a complex process where each conclusion must be supported by evidence rigorously collected, giving it full credibility.

In this paper, we first address these problems by proposing an approach to reconstruct scenarios from suspect data and analyse them using semantic tools and knowledge from experts. Secondly, this paper answers the challenge of correctness of the whole investigative process with a formal incident modelization and timeline reconstruction and analysis operators. The paper is organized as follows. Section Related Works reviews important issues of the events reconstruction problem and the various approaches proposed so far. The SADFC (*Semantic Analysis of Digital Forensic Cases*) approach is described in Section SADFC approach, and the formal advanced timeline reconstruction and analysis model is presented in Section Advanced timeline analysis model. Finally, a case study illustrating the key characteristics of the proposed approach is given in Section Case study.

**Related works**

Events reconstruction has many issues, which are directly related to the size of the data, digital investigation process complexity, and IT infrastructures challenges. For instance, Table 1 compares some existing approaches (their strengths (✓), limitations (✘), partial or inadequate solutions (●)) with regard to some key issues, such as heterogeneity, automatic knowledge extraction, the use of proven theory as support, analysis capabilities, and preservation of data integrity. While some of these challenges have been a focus for many researchers and developers for the last decade, the size of data volumes (Richard III & Roussev, 2006) and data heterogeneity are still very challenging. The first (large data sizes) introduces many challenges at every phase of the investigation process; from the data collection to the interpretation of the results. The second (data heterogeneity) is usually due to multiple footprint sources such as log files, information contained in file systems, etc. We can classify events heterogeneity into three categories:

- *Format*: The information encoding is not the same among sources due to the formatting. Therefore, depending on the source, footprint data may be different.
- *Temporal*: The use of different sources from different machines may cause timing problems. First, there are some issues due to the use of different time zones and unsynchronized clocks. Second, the temporal heterogeneity can be due to the use of different formats or granularities (e.g. 2 s in FAT file systems, 100 ns in NTFS file systems).
- *Semantic*: The same event can be interpreted or represented in different ways. For example, an event describing the visit of a webpage may appear in different ways in web browser logs and server logs.

In order to gather all the events found in footprint sources in a single timeline, a good handling of all these forms of heterogeneity is required. This leads to the development of an automated information processing approach which is able to extract knowledge from these heterogeneous sources. In addition, once extracted, this knowledge should be federated within the same model so as to facilitate their interpretation and future analysis. The effectiveness of a such approach can be assessed by the following criteria:

- Efficient automated tools that can extract events and build a timeline (Criterion 1 in Table 1).
- The ability to process multiple and various footprint sources and to federate the information collected in a coherent and structured way (Criterion 2 in Table 1).
- The ability to assist users during the timeline analysis. This latter encompasses many aspects such as making the timeline easier to read, identifying correlations between events or producing conclusions from knowledge contained in the timeline (Criterion 3 in Table 1).

For the majority of existing approaches, solutions are provided to automatically extract events and construct the timeline. Chen et al. (2003) introduced a set of automated extractors to collect events and store them in a canonical database, which allows to quickly generate a temporal ordered sequence of events. These automatic extractors, a widely used concept, can also generate the timeline (Olsson and Boldt, 2009; Gudhjonsson, 2010; Hargreaves and Patterson, 2012). However, current tools extract data in its raw form without a good understanding of the meaning of footprints, which makes their analysis more difficult. In order to deal with the semantic heterogeneity, the FORE (Forensics of Rich Events) system stores the events in an ontology (Schatz et al., 2004a,b). This ontology uses the notions of entity and event to represent the state change of an object over time. Nevertheless, the time model implemented in this ontology is not accurate enough (use of instant rather than interval) to represent events accurately. In addition, the semantic coverage of this ontology can be improved. Another storage structure is also proposed in Buchholz and Falk (2005). They proposed a scalable structure, a variant of a balanced binary search tree. However, one of the problems of this approach is that one has to build the case by manually selecting relevant events.

**Table 1**
Evaluation of approaches.

| Approach/Criterion | Auto extraction | Heterogeneity | Analysis | Theory | Data integrity |
|---|---|---|---|---|---|
| ECF (Chen et al., 2003) | ✓ | ✓ | ✘ | ✘ | ✘ |
| FORE (Schatz et al., 2004b) | ✓ | ✓ | ● | ✘ | ✘ |
| Finite state machine(Gladyshev and Patel, 2004) | ✘ | ● | ● | ✓ | ✘ |
| Zeitline (Buchholz and Falk, 2005) | ● | ✓ | ✘ | ✘ | ✓ |
| Neural networks(Khan and Wakeman, 2006) | ● | ● | ✘ | ● | ✘ |
| CyberForensic TimeLab(Olsson and Boldt, 2009) | ✓ | ✓ | ✘ | ✘ | ✘ |
| log2timeline (Gudhjonsson, 2010) | ✓ | ✓ | ✘ | ✘ | ✘ |
| Timeline reconstruction (Hargreaves and Patterson, 2012) | ✓ | ✓ | ● | ● | ✘ |

Some automatic events extraction approaches are also able to process heterogeneous sources. This usually consists of creating a set of extractors dedicated to each type of footprint sources. In Gudhjonsson (2010), the authors highlight several limitations of the current event reconstruction systems such as the use of a small number of event sources, which makes the timeline vulnerable to anti-forensics techniques (e.g., alteration of timestamps). They proposed to use a large number of event sources to ensure that the high quality of the timeline and the impact of anti-forensics techniques is minimized. *log2timeline* uses various sources including Windows logs, history of various web browsers, log files of other software resources, etc. Multiple heterogeneous sources require a consistent representation of the events in order to process them. The majority of approaches propose their own model for event representation. The most commonly used event features include the date and time of the event (an instant or an interval when the duration of events is included) and information about the nature of the event.

Regarding the timeline analysis, few solutions have been proposed. The FORE approach attempts to identify event correlations by connecting events with causal or effect links. In Gladyshev and Patel (2004), the authors try to perform the event reconstruction by representing the system behaviour as a finite state machine. The event reconstruction process can be seen as a search for sequences of transitions that satisfy the constraints imposed by the evidence. Then, some scenarios are removed using the evidence collected (thus, there is no automation of the extraction process in this approach). In Hargreaves and Patterson (2012), they proposed a pattern-based process to produce high-level events ("human-understandable" events) from a timeline containing low-level events (events directly extracted from the sources). Although the proposed approach is relevant, it can handle only one of the many aspects of the analysis by helping the investigator to read the timeline much easier. Therefore, other aspects such as causality analysis between events are not covered in this approach.

All approaches have to satisfy some key requirements such as credibility, integrity, and reproducibility of the digital evidence (Baryamureeba and Tushabe, 2004). In recent years, the protagonists of digital forensics moved away from investigative techniques that are based on the investigators experience, to techniques based on proven theories. It is also necessary to provide clear explanation about the evidence found. In addition, one has to ensure that the tools used do not modify the data collected on crime scenes. Thus, it is necessary to develop tools that extract evidence, while preserving the integrity of the data. Finally, a formal and standard definition of the reconstruction process is needed to ensure the reproducibility of the investigation process and credibility of the results. A clearly-defined investigation model allows to explain the process used to get the results. To this end we believe that the following criteria are crucial for such techniques: the use of a theoretical model to support the proposed approach (criterion 4 in Table 1) and the ability to maintain the data integrity (criterion 5 in Table 1). As a prelude to his work, Gladyshev (Gladyshev and Patel, 2004) argued that a formalization of the event reconstruction problem is needed to simplify the automation of the process and to ensure the completeness of the reconstruction. In Khan and Wakeman (2006), the authors proposed an event reconstruction system based on neural networks. The use of a machine learning technique appears to be a suitable solution because it is possible to know the assumptions and reasoning used to obtain final results. However, neural networks behaviour is not entirely clear (especially during the training step). Thus, the approach adopted by Khan does not seem to fulfil the goal of making reasoning explicit. In Hargreaves and Patterson (2012), the system keeps information about the analysis which leads to infer each high-level event. This gives the opportunity to provide further information when needed. As for the preservation of the data integrity, the approach described in Buchholz and Falk (2005) uses a set of restrictions to prevent the alteration of the evidence. Regarding the process model, numerous investigation models have been proposed, however, none of them is designed for automated or semi-automated investigation. Existing models (DFWRS model (Palmer, 2001), End to End Digital Investigation (Stephenson, 2003), Event-Based Digital Forensics Investigation Framework (Carrier and Spafford, 2004a), Enhanced Digital Investigation Process Model (Baryamureeba and Tushabe, 2004), Extended Model of Cybercrime Investigation (Ciardhuáin, 2004), Framework for a Digital Forensics Investigation (Kohn et al., 2006)) are designed to guide humans investigators by providing a list of tasks to perform. Thus, the proposed models are not accurate enough to provide a framework for the development of automated investigation tools. Among the limitations of the existing frameworks, the characterization of the data flow through the model is absent and the meaning of the steps is not clear. The creation of a well-defined and explicit framework is needed to allow an easy translation of the investigation process into algorithms.

We can see in Table 1 that existing approaches are not able to fulfil all the criteria. Two limitations of the existing approaches are the lack of automation of the timeline analysis and the absence of theoretical foundations to explain the conclusions produced by the tools. The assistance provided to the user should not be limited only to the construction of the timeline. It is necessary to deal with a large amount of events (criterion 3), the heterogeneity of event sources, the need to federate events in a suitable model (see criterion 1 and criterion 2), data integrity, and the whole investigative process correctness and validation. The purpose of this study is to propose an approach that satisfies all the criteria presented in Table 1 while producing all the necessary evidence in an efficient way.

## SADFC approach

To reach these objectives, we propose the following approach:

- Identify and model the knowledge related to an incident and the knowledge related to the investigation process used to determine the circumstances of the incident:
  - Introduce a formalization of the event reconstruction problem by formally defining entities involved in an incident.
  - Define a knowledge representation model based on the previous definitions used to store knowledge about an incident and knowledge used to solve the case (to give credibility to the results and to ensure the reproducibility of the process).
- Provide extraction methods of the knowledge contained in heterogeneous sources to populate the knowledge model.
- Provide tools to assist investigators in the analysis of the knowledge extracted from the incident.

Unlike conventional approaches, the SADFC approach uses techniques from knowledge management, semantic web and data mining at various phases of the investigation process. Issues related to knowledge management and ontology are two central elements of this approach as they deal with a large part of the mentioned challenges. According to Gruber (1993), "an ontology is an explicit, formal specification of a shared conceptualization". Thus, ontology allows to represent knowledge generated during an investigation (knowledge about footprints, events, objects etc.). Ontology provides several advantages such as the possibility to use automatic processes to reason on knowledge thanks to its formal and explicit nature, the availability of rich semantics to represent knowledge (richer than databases due to its sophisticated semantic concepts (Martinez-Cruz et al., 2012)) and the possibility of building a common vision of a topic that can be shared by investigators and software developers. Ontology has already proved its relevance in computer forensics (Schatz et al., 2004a,b). The use of ontology is also motivated by its successful use in other fields such as biology (Schulze-Kremer, 1998) and life-cycle management (Vanlande et al., 2008).

SADFC is a synergy of the three elements presented below, which, once assembled, constitute a coherent package describing methods, processes and technological solutions needed for events reconstruction:

- *Knowledge Model for Advanced Digital Forensics Timeline Analysis* which is presented in the following part of the paper.
- *Investigation Process Model*: The aim of this model is to define the various phases of the event reconstruction process: their types, the order between them and the data flow through the whole process.
- *Ontology-centred architecture*: This architecture consists of several modules which implement some of the key functions such as footprint extraction, knowledge management, ontology and the visualization of the final timeline. This architecture is based on an ontology which implements the knowledge model proposed in our approach. As the proposed ontologies in Schatz et al. (2004a,b) are not advanced enough, we have designed and developed a new ontology. For instance, some relations included in our knowledge model are absent in the ontology of the FORE system.

In the following sections, we present a formalized knowledge model for advanced forensics timeline analysis, while the investigation process model and the architecture will be discussed in details in other documents.

## Advanced timeline analysis model

This section describes the knowledge model used in the SADFC approach to allow to perform an in-depth analysis of timeline while fulfilling the law requirements. Models proposed in the literature (Schatz et al., 2004a,b) are still limited in terms of the amount of knowledge they can store about events and therefore, analysis capabilities. Temporal characteristics predominate over other aspects such as interaction of events with objects, processes or people. Knowledge representing the relations between events and others entities is not sufficiently diversified for the subsequent analysis required. Thus, we propose a rich knowledge representation containing a large set of entities and relations and allowing to build automated analysis processes. In addition, the proposed model is designed to meet the legal requirements and contains knowledge allowing to reproduce the investigative process and to give full credibility to the results. It should be noted that the assumption that the data has been processed upstream to ensure accuracy and correctness is used. A digital forensic investigation process model including processes for consistency check of data and filtering will be proposed in future works.

We will first formally define the entities and relations composing the model and then, introduce a set of operators to manipulate the knowledge. An overview of the proposed knowledge model is given in Fig. 1.

*Formal modelling of incident*

We first define entities of our knowledge model and then we detail the four composed relations using this knowledge.

*Subject, object, event and footprint*
A crime scene is a space where a set of events $E = \{e_1, e_2, \ldots, e_i\}$ takes place. An event is a single action occurring at a given time and lasting a certain duration. An event may be the drafting of a document, the reading of a webpage or a conversation via instant messaging software.

*Subject.* During its life cycle, an event involves *subjects*. Let $S$ be the set containing subjects covering human actors and processes (e.g. Firefox web browser, Windows operating system, etc.), a subject $x \in S$ corresponds to an entity involved in one or more events $e \in E$ and is defined by $x = \{a \in A_s | x\, \alpha_s\, a\}$ where:

- $A_s$ is a set containing all the attributes which can be used to describe a subject. A subject attribute may be the first name and the last name of a person, the identifier of a web session, the name of a Windows session, etc.

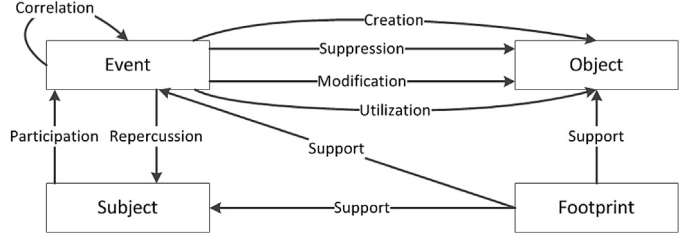

Fig. 1. Knowledge model.

- $\alpha_s$ is the relation used to link a subject with the attributes of $A_s$ describing it.

*Object.* During its life cycle, an event can also interact with *objects*. An object may be a webpage, a file or a registry key for example. An object $x \in O$ is defined by $x = \{a \in A_o \mid x \alpha_o a\}$ where:

- $A_o$ is a set containing all the attributes which can be used to describe an object. An object attribute may be for example a filename, the location of an object on a hard disk, etc.
- $\alpha_o$ is the relation used to link an object with the attributes of $A_o$ describing it.
- $O \subseteq \wp(A_o)$ is the set of the objects, meaning that $o \in O$ belongs to the power set of $A_o$. An object is a composition of one or several attributes of $A_o$.

Note, that for easier human understanding, an object can also be seen as a composition of attributes and objects (because objects are sets of attributes), e.g. a registry key is an object composed of several attributes such as its value and its key name. This registry key is also an attribute of the object representing the database containing all keys of the system.

*Event.* Each event takes place in a time-interval defined by a start time and an end time. These boundaries define the life cycle of the event. The use of a time interval allows to represent the notion of uncertainty (Liebig et al., 1999). For example, when the start time of an event cannot be determined accurately, the use of a time-interval allows to approximate it. The use of intervals requires the introduction of a specific algebra (e.g. to order events). In our work, the Allen algebra, illustrated in the Table 2, is used (Allen, 1983). The authors of this paper are aware of the problems caused by temporal heterogeneity and anti-forensics techniques on the quality and the accuracy of timestamps (granularity, timestamps offset and alteration, timezone, etc). Several works try to characterize the phenomena related to the use of time information in digital forensic investigations and to propose techniques to solve these problems (Schatz et al., 2006; Gladyshev and Patel, 2005; Forte, 2004). In this paper, we assume that all timestamps used in our model are adjusted and normalized beforehand by a process that will be the subject of future work (the proposed model is generic enough to incorporate future solutions). We consider that each function returns the value 1 if events meet the constraints and 0 otherwise.

An *event* $e \in E$ is defined by $e = \{t_{start}, t_{end}, l, S_e, O_e, E_e\}$ where:

- $t_{start}$ is the start time of the event, $t_{end}$ is the end-time of the event and $l$ is the location where the event took place. This location may be a machine (represented by an IP) and its geolocalisation.
- $S_e$ is a set containing all subjects involved in the event. $S_e = \{s \in S \mid e \in E, s \sigma_s e\}$ where $\sigma_s$ is a composed relation used to link an event $e \in E$ with a subject $s \in S$. The relation $\sigma_s$ is defined below.
- $O_e$ is a set containing all objects related to the event $e$. $O_e = \{o \in O \mid e \in E, e \sigma_o o\}$ where $\sigma_o$ is a composed relation used to link an event $e \in E$ with an object $o \in O$. The relation $\sigma_o$ is defined below.
- $E_e$ is the set containing all events with which the event is correlated. $E_e = \{x \in E \mid e \in E, e \sigma_e x\}$ where $\sigma_e$ is a composed relation used to link an event $e \in E$ with an event $x \in E$. The relation $\sigma_e$ is defined below.

*Footprint.* According to Ribaux (2013), a *footprint* is the sign of a past activity and a piece of information allowing to reconstruct past events. A footprint may be a log entry or a web history for example as a log entry gives information about software activities and web histories provide information about user's behaviour on the Web. Let $F$ be a set containing all footprints related to a case, a footprint $x \in F$ is defined by $x = \{a \in A_f \mid x \alpha_f a\}$ where:

Table 2
Allen algebra.

| Functions | Constraints | Example |
|---|---|---|
| before(X,Y) | $x_{t_{end}} < y_{t_{start}}$ | X —— Y |
| equal(X,Y) | $x_{t_{start}} = y_{t_{start}}$ && $x_{t_{end}} = y_{t_{end}}$ | X / Y |
| meets(X,Y) | $x_{t_{end}} = y_{t_{start}}$ | X Y |
| overlaps(X,Y) | $x_{t_{start}} < y_{t_{start}}$ && $x_{t_{end}} > y_{t_{start}}$ | X Y |
| during(X,Y) | $x_{t_{start}} > y_{t_{start}}$ && $x_{t_{end}} < y_{t_{end}}$ | X Y |
| starts(X,Y) | $x_{t_{start}} = y_{t_{start}}$ | X Y |
| finishes(X,Y) | $x_{t_{end}} = y_{t_{end}}$ | X Y |

- $A_f$ is a set containing all the attributes which can be used to describe a footprint. For example, a bookmark contains attributes such as the title of the bookmark, the date of creation and the webpage pointed by the bookmark.
- $\alpha_f$ is the relation used to link a footprint with the attributes of $A_f$ used to described it.

Footprints are the only available information to define past events and can be used by investigators to reconstruct the events which happened during an incident. However, the imperfect and incomplete nature of the footprints can lead to produce approximate results. It is therefore not always possible to determine which event is associated with a given footprint. In addition, it is not always possible to fully reconstruct an event from a footprint. Thus, a footprint can be used to identify one or several features:

- The temporal features of an event. For example, a footprint extracted from the table moz_formhistory (database FormHistory of the web browser Firefox) can be used to establish the time at which a form field has been filled.
- A relation between an event and an object. For example, a footprint extracted from the table moz_historyvisits (database Places of the web browser Firefox) can be used to link an event representing the visit of a webpage to this webpage.
- A relation between an event and a subject. For example, all footprints produced by the web browser Firefox are stored in a folder named with the profile name of the user. This allows to link each Firefox event to the user designated by this name.
- The features of an object. For example, a footprint extracted from the table moz_places (database Places) can be used to determine the URL and the title of a webpage.
- The features of a subject. For example, a footprint extracted from the table moz_historyvisits (database Places) can be used to determine the session identifier of a user.

*Relations*

To link the entities presented before, four composed relations which were introduced in previous part are detailed here.

*Subjects relations.* $\sigma_s$ is composed of two types of relations to link an event $e \in E$ with a subject $s \in S$ and can be defined in the following way $\sigma_s = s\ isInvolved\ e \vee s\ undergoes\ e$:

- *Relation of participation*: $s\ isInvolved\ e$ means that $s$ initiated or was involved in $e$. For example, the user of a computer is involved in an event representing the login to the session, etc.
- *Relation of repercussion*: $s\ undergoes\ e$ means that $s$ is affected by the execution of $e$. For example, a user is affected by the removal of one of his files, etc.

*Objects relations.* $\sigma_o$ is composed of four types of relations to link an event $e \in E$ with an object $o \in O$ and can be defined in the following way $\sigma_o = e\ creates\ o \vee e\ removes\ o \vee e\ modifies\ o \vee e\ uses\ o$:

- *Relation of creation*: $e\ creates\ o$ means that $o$ does not exist before the execution of $e$ and that $o$ is created by $e$.
- *Relation of suppression*: $e\ removes\ o$ means that $o$ does not exist anymore after the execution of $e$ and that $o$ is deleted by $e$.
- *Relation of modification*: $e\ modifies\ o$ means that one or more attributes of $o$ are modified during the execution of $e$.
- *Relation of usage*: $e\ uses\ o$ means that one or more attributes of $o$ are used by $e$ to carry out its task.

*Events relations.* $\sigma_e$ is composed of relations used to link two events $x, e \in E$ and can be defined in the following way $\sigma_e = x\ composes\ e \vee e\ composes\ x \vee x\ causes\ e \vee e\ causes\ x$. In our works, $x\ is\ Correlated\ e$ means that $x$ is linked to $e$ on the basis of multiple criteria: use of common resources, participation of a common person or process, temporal position of events. We distinguish two special cases of the relation of correlation:

- *Relation of composition*: $x\ composes\ e$ means that $x$ is an event composing $e$. For example, an event representing a Windows session is composed of all events initiated by the user during this session. Let $x = \{t_{xstart}, t_{xend}, S_x, O_x, E_x\}$ be an event composing $e = \{t_{estart}, t_{eend}, S_e, O_e, E_e\}$, the relation of composition implies a set of constraints. First, a temporal constraint requiring that sub-events take place during the parent event. Using Allen relations, if $x\ composes\ e$ then $equal(x,e)$ or $during(x,e)$ or $starts(x,e)$ or $finishes(x,e)$. Sub-events have also constraints on participating subjects as well as the objects with which the event interacts. If $x\ composes\ e$ then $S_x \subseteq S_e$ and $O_x \subseteq O_e$. Thus, $x\ composes\ e = [equal(x,e) \vee during(x,e) \vee starts(x,e) \vee finishes(x,e)] \wedge (S_x \subseteq S_e) \wedge (O_x \subseteq O_e)$.
- *Relation of causality*: $x\ causes\ e$ means that $x$ has to happen to allow $e$ to happen. For example, an event describing the download of a file from a server is caused by the event describing the connection to this server. An event can have several causes and can be the cause of several events. Let $e = \{t_{estart}, t_{eend}, S_e, O_e, E_e\}$ be an event caused by $x = \{t_{xstart}, t_{xend}, S_x, O_x, E_x\}$, the relation of causality implies a temporal constraint requiring that the cause must happen before the consequence. Using Allen algebra, $x\ causes\ e = [before(x,e) \vee meets(x,e) \vee overlaps(x,e) \vee starts(x,e)] \vee (S_x \cap S_e) \vee (O_x \cap O_e)$.

*Footprints relation.* $\sigma_f$ is a relation used to link a footprint $f \in F$ with an entity $en \in \{E \times O \times S\}$. This relation is called *Relation of support*: $f\ supports\ en$ means that $f$ is used to deduce one or more attributes of $en$. We define a function *support* which can be used to know the footprints used to deduce a given entity: $support(en \in \{E \times O \times S\}) = \{f \in F | f \sigma_f en\}$. For example, an entry of a web history can be used to reconstruct an event representing the visit of a webpage by a user.

*Crime scene*

In our works, we define a *crime scene CS* as an environment in which an incident takes place and by $CS = \{PCS, DCS\}$. *PCS* is a set containing the *physical crime*

*scenes*. At the beginning of an investigation, *PCS* is initialized with the localization where the incident takes place. However, in an investigation, the crime scene is not limited to only one building. Due to network communication for example, the initial physical crime scene may be extended to a set of new physical crime scenes if one of the protagonists communicated with an other person through the network, downloaded a file from a remote server, etc. In these cases, the seizure of the remote machines (and by extension, the creation of new physical crime scenes and digital crime scenes) should be taken into account as it may be relevant for the investigation. *DCS* is a set containing the *digital crime scenes*. Unlike Carrier (Carrier and Spafford, 2003), there is no distinction between primary and secondary digital crime scene in our works to simplify the modelization.

A crime scene is also related to events that take place on it $E_{CS} = \{E_{iCS} \cup E_{cCS} \cup E_{nCS}\}$. For easier notation, we write here $E = \{E_i \cup E_c \cup E_n\}$ where:

- $E_i$ is a set containing *illicit* events as $E_i = \{e \in E \mid e\, \sigma_i\, i, i \in I\}$ with $I$ a set containing all actions considered as infractions by the laws and $\sigma_i$ the relation linking an event with an action. For example, an event representing the upload of defamatory documents to a website is an event of $E_i$.
- $E_c$ is a set containing the *correlated* events as $E_c = \{e \in E \mid e\, \sigma_i\, l, l \in L, e\, \sigma_e\, x, x \in E_i\}$. This set contains all legal events which are linked with a set of illicit events $x$.
- $E_n = \{E \setminus (E_i \cup E_c)\}$ is a set containing the events which are not relevant for the investigation.

*Event reconstruction and analysis operators*

Since investigators have ensure the preservation of the crime scene, it becomes a protected static environment containing a set of footprints. After the collection of all the footprints of the crime scene, the goal of the event reconstruction consists in moving from the static crime scene to a timeline describing the dynamics of an incident which happened in the past. Describing an incident means identify all the events $E_{inc} = E_i \cup E_c$ using footprints of *F*. Events ordered chronologically describing an incident are called the *scenario* of the incident. In the SADFC approach, four operators are defined to carry out the event reconstruction. These operators are illustrated in the Section Case study.

The aim of *extraction operators* is to identify and extract relevant information contained in digital footprints from various sources. Sources are chosen according to the definition of the perimeter of the crime scene. The relevancy or the irrelevancy of information contained in footprints is determined by the investigators according to the goals of the investigation. For example, in a case involving illegal downloads of files, the investigator will pay more attention to information related to the user behaviour on the web while logs of word processing software will be ignored. The *mapping operators* create entities (events, objects and subjects) associated to the extracted footprints. These operators take the form of mapping rules allowing to connect attributes extracted from footprints and attributes of events, objects and subjects. A large part of the features of an event can be determined by extraction operators from footprints collected in the crime scene. However, the identification of some kinds of features requires the use of advanced techniques such as inference. The *inference operators* allow to deduce new knowledge about entities from existing knowledge. Unlike extraction operators which use knowledge of footprints, inference operators use the knowledge about events, objects and subjects (knowledge generated by mapping operators). The *analysis operators* are used to help the investigators during the interpretation of the timeline. These operators can be used to identify relations between events or to highlight the relevant information of the timeline. In our works, we introduce an operator dedicated to the identification of event correlations. The correlation between two events $e, x \in E$ is measured by the following function:

$$\text{Correlation}(e,x) = \text{Correlation}_T(e,x) + \text{Correlation}_S(e,x)$$
$$+ \text{Correlation}_O(e,x) + \text{Correlation}_{KBR}(e,x)$$
(1)

*Correlation*($e,x$) can be weighted to allow to give more importance to one of the correlation functions. *Correlation*($e,x$) can also be ordered and thresholded to deal with data volume constraints by selecting the most significant correlations. Those four correlations are described in the following way:

*Temporal Correlation, Correlation$_T$($e,x$)*. First of all, a set of assumptions about the temporal aspect is defined (according to the Allen algebra given in Table 2):

- The greater the relative difference between the two events *before*($e,x$) is, the lower the temporal relatedness is and reciprocally.
- The temporal relatedness is high for functions *meets*($e,x$), *overlaps*($e,x$), *during*($e,x$) and *finishes*($e,x$) and maximal for *starts*($e,x$) and *equals*($e,x$).

Thus, $\text{Correlation}_T(e,x) = \alpha * \text{starts}(e,x) + \alpha * \text{equals}(e,x)$
$$+ \text{meets}(e,x) + \text{overlaps}(e,x)$$
$$+ \text{during}(e,x) + \text{finishes}(e,x)$$
$$+ \text{before}(e,x)$$
(2)

where *starts*($e,x$), *equals*($e,x$), *meets*($e,x$), *overlaps*($e,x$), *during*($e,x$), *finishes*($e,x$) are binary functions and *before*($e,x$) $= 1/(x_{tstart} - e_{tend})$. Previous assumptions state that the more two events are close in time, the more it is likely that these events are correlated. Because of time granularities, and multi-tasks computers, if two events start at the same time, the relatedness is more important. That is why this importance can be highlighted with an $\alpha$ factor.

*Subject Correlation, Correlation$_S$($e,x$) and Object Correlation, Correlation$_O$($e,x$)*. These two correlations respectively quantify correlations regarding subjects involved in each event and objects used, generated, modified or removed by events. The following hypothesis are defined according to the core idea of several domains such as *data mining*, for example in formal concept analysis (Ganter et al., 1997) which groups objects regarding to the attributes they share or in *statistics* such as principal component analysis, which groups observations measuring how far they are spread (variance).

- The relatedness between *e* and *x* increases proportionally to the number of common subjects they share regarding to relations of *participation* and *repercussion*. $Correlation_S(e,x) = |S_e \cap S_x| / max(|S_e|,|S_x|)$.
- The relatedness between *e* and *x* increases proportionally to the number of objects they share regarding to relations of *creation*, *suppression*, *modification* and *usage*. $Correlation_O(e,x) = |O_e \cap O_x| / max(|O_e|,|O_x|)$.

*Rule-based Correlation*, $Correlation_{KBR}(e,x)$. In addition to the previous factors (time, subject, object), rules based on expert knowledge can be used to correlate events. $Correlation_{KBR}(e,x) = \sum_{r=1}^{n} rule_r(e,x)$ with $rule_r(e,x) = 1$ if the rule is satisfied and 0 otherwise. Therefore, the greater the number of satisfied rules is, the greater the value of $Correlation_{KBR}$ is. All correlation functions return a score between 0 and 1 except $Correlation_{KBR}$. This specificity allows to give more importance to expert knowledge synthesized in the rules.

**Case study**

The aim of this section is to illustrate with an example how this knowledge model can be used to formalize a computer forensics case and reconstruct the incident by using the proposed operators. To illustrate the capabilities of our model, we designed a fictitious investigation concerning a company manager who contacted a private investigator after suspicions about one of its employees. The manager believes that this employee uses the internet connection of the company to illegally download files for personal purposes. As the investigation process model is not presented yet, we used in this case study a process designed for the need of this paper and composed of five steps: the definition of the crime scene, the collection of footprints, the creation of entities (event, object, subject), the enrichment of the extracted knowledge and finally the construction and the analysis of the timeline.

*Crime scene definition*

As a start of the investigation, the investigator evaluates the size of the crime scene. As suspicions concern the workstation of the employee, the open-space of the company where this computer is located is designated as the physical crime scene. Based on the testimony and to reduce the complexity of the investigation, only the workstation of the suspected employee is added to the *DCS* set. To establish whether or not the doubts on the employee are founded, the investigator in charge of the case seizes the machine used by the employee and starts using the theoretical tools proposed in this paper to track the user's past activities.

*Footprints collection*

After defining the machine used by the employee as crime scene, the investigator collects footprints on this latter. To carry out this step, the extraction operators are used. According to the objectives of the investigation, the investigator chooses to collect footprints left by the web browser used by the employee as he knows that he can find information about downloads performed by the user. The output of the extraction is a set containing several types of web browser footprints (*id* is a unique identifier for each element populating our model):

- *fWebpage*(*id,webpageID,pageTitle,URL,hostname*): footprint giving information about a webpage.
- *fVisit*(*id,date,session,pageID*): footprint of a visit of a webpage.
- *fBookmark*(*id,date,bookmarkTitle,pageID*): footprint of a creation of a bookmark.
- *fDownload*(*id,start,end,filename,pageSourceID,URLTarget*): footprint of a file download where *start* and *end* are the dates on which the download starts and ends, *pageSourceID* is the webpage from where the download is launched and *URLTarget* is the destination used to store the file.

The result of the extraction is given in Fig. 3 and is illustrated in the left part of Fig. 2.

*Entities creation*

The output of this step is a set containing the entities (event, object, subject) which can be recovered using the

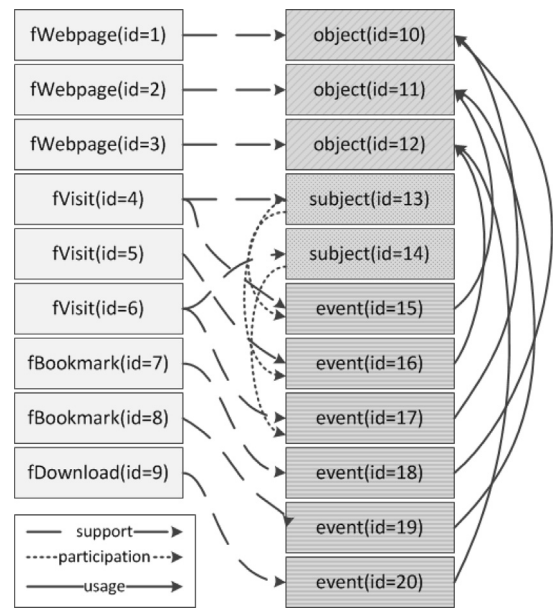

**Fig. 2.** Overview of the knowledge generated during the case.

```
fWebpage(1, 153, "BBC_News_Home",
    "http://www.bbc.com/news/", "www.bbc.com").
fWebpage(2, 165, "Torrent,_Streaming,_Crack,_Serial...",
    "http://www.warez.com", "www.warez.com").
fWebpage(3, 28, "Warez:_Films",
    "http://www.warez.com/films", "www.warez.com").
fVisit(4,"2013-08-14T10:35:43",351,165).
fVisit(5,"2013-08-14T10:37:02",351,28).
fVisit(6,"2013-08-14T10:55:41",410,153).
fBookmark(7,"2013-08-14T10:55:59","News",153).
fBookmark(8,"2013-08-14T10:35:53","Download",165).
fDownload(9,"2013-08-14T10:37:20","2013-08-14T11:22:12",
"changingSeasons.divx", 28, "C:\Users\UserA\Desktop\").
```

**Fig. 3.** Output of footprints extraction.

information left in the digital crime scene. To carry out this step, the mapping operators are used. The output of the mapping is the set of entities given in Fig. 4 and illustrated in the right part of Fig. 2. Entities are linked together by relations defined in Section Relations. $support(x,y)$ means that the footprint identified by the id $x$ has been used to create the entity identify by the id $y$. $participation(x,y)$ means that the subject $x$ is involved in the event $y$. $usage(x,y)$ means that the event $x$ used the object $y$. A subject is defined by $subject(id,session)$ where $id$ is a unique identifier while $session$ is the session number associated to the user. The hypothesis that duration of bookmark creation and the visit of a webpage is null is used (start time equals to end time).

### Knowledge enrichment

This step is particularly useful to improve the results of the analysis steps as it allows to discover new knowledge about entities. For example, the only available information to determine the subject involved in an event generated by a web browser is the session identifier found in some digital footprints extracted from Web browser. To identify the subject involved in other events, we used an inference operator based on the following assumption. Let $e_i$ be the first visit of a webpage for the session $s$, $e_j$ the last visit of $s$, $t_i$ the start date of $e_i$ and $t_j$ the end date of $e_j$, an event occurring on the machine in a date within the time interval defined by $t_i$ and $t_j$ involves the person who owns the session $s$. In this case study, the person who is involved in the event $event_{id=19}(j)$ in Fig. 4 (creation of a bookmark for the webpage "http://www.warez.com") is unknown. However, using the previous inference rule, it is possible to infer that the user is the subject $subject_{id=13}(d)$ using the fact that this user perform a visit of a webpage before and after the event $event_{id=19}(j)$ (respectively $event_{id=15}(f)$ and $event_{id=16}(g)$). Thus, the fact $participation(13,19)$ can be added.

### Timeline construction and analysis

After collecting knowledge about events and related entities, the last step is to build the associated timeline and analyse it. A graphical representation of the timeline is given in Fig. 5. To find correlations between events, the operator introduce in Section Event reconstruction and

```
object(10, 153, "BBC_News_Home",
    "http://www.bbc.com/news/", "www.bbc.com"). (a)
object(11, 165, "Torrent,_Streaming,_Crack,_Serial...",
    "http://www.warez.com", "www.warez.com"). (b)
object(12, 28, "Warez:_Films",
    "http://www.warez.com/films", "www.warez.com"). (c)
support(1,10).   support(2,11).   support(3,12).
support(4,13).   subject(13, 351). (d)
support(6,14).   subject(14, 410). (e)
event(15, "2013-08-14T10:35:43", "2013-08-14T10:35:43",
    "153.168.1.1"). (f)
support(4,15).   usage(15,11).   participation(13,15).
event(16, "2013-08-14T10:37:02", "2013-08-14T10:37:02",
    "153.168.1.1"). (g)
support(5,16).   usage(16,12).   participation(13,16).
event(17, "2013-08-14T10:55:41", "2013-08-14T10:55:41",
    "153.168.1.1"). (h)
support(6,17).   usage(17,10).   participation(14,17).
event(18, "2013-08-14T10:55:59", "2013-08-14T10:55:59",
    "153.168.1.1"). (i)
event(19, "2013-08-14T10:35:53", "2013-08-14T10:35:53",
    "153.168.1.1"). (j)
event(20, "2013-08-14T10:37:20", "2013-08-14T11:22:12",
    "153.168.1.1"). (k)
support(7,18).   usage(18,10).   support(8,19).
    usage(19,11).   support(9,20).   usage(20,12).
```

**Fig. 4.** Entities created by the mapping process.

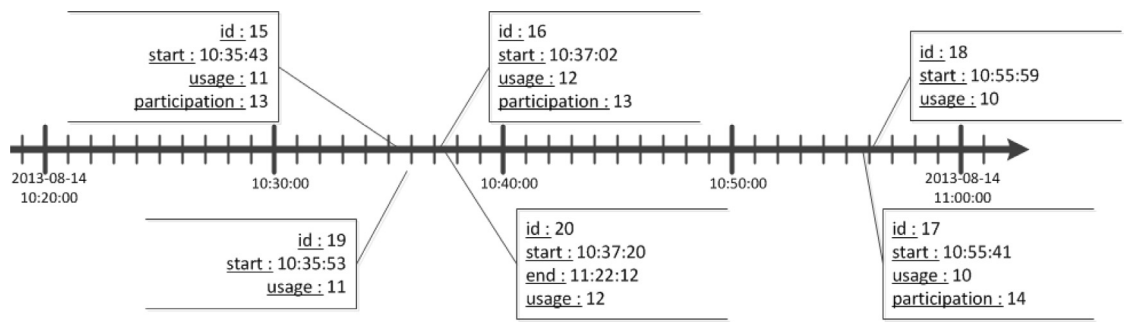

**Fig. 5.** Timeline of the incident.

analysis operators is used. To illustrate the computation of correlations, three events defined in Fig. 4 are used. In this example, the correlations between the events $event_{id=15}(f)$ and $event_{id=17}(h)$ (called "pair A") and between the events $event_{id=15}(f)$ and $event_{id=19}(j)$ (called "pair B") are computed. First, the subject correlation and the object correlation are computed. Regarding the pair A, the two events share neither subjects nor objects. Regarding the pair B, the two events use a common resource which is the webpage "http://www.warez.com" ($object_{id=11}(b)$). A common subject is also involved in both events: the subject identified by the session number 14 ($subject_{id=13}(d)$). Second, the temporal correlation is computed. For both pairs, the case $before(x,e)$ where an event occurs before the other is used. Thus, the temporal correlation is greater for the pair B than the pair A as the relative difference between $event_{id=15}(f)$ and $event_{id=19}(j)$ is lower than the difference between $event_{id=15}(f)$ and $event_{id=17}(h)$. To conclude, the pair B is more correlated than the pair A due to the use of a common object ($object_{id=11}(b)$), the participation of a common subject ($subject_{id=13}(d)$) and the temporal proximity between these two events. At the end of the analysis step, the investigator gets a timeline enriched with information about correlations between events. The investigator starts the interpretation of the results by identifying $event_{id=15}(f)$ which is a visit of the website "warez.com" that appears to be a web platform providing links to illegal files. Then, correlation results show that $event_{id=15}(f)$ and $event_{id=19}(j)$ are highly correlated. This correlation allows to conclude that the visit of the website is voluntary and that this website has some relevance for the user (indeed, he uses a bookmark to facilitate a subsequent visit). The analysis of others events and others possible correlations can lead to the conclusion than the user has visited intentionally the website "warez.com" and used this website to download illegal files.

## Conclusion and future works

In digital investigations, one of the most important challenge is the reconstruction and the analysis of past events, because of the heterogeneity of data and the volume of data to process. To answer this challenge we present the SADFC approach. It allows to reduce the tedious character of the analysis thanks to automatic analysis processes helpful for the investigators. Thus, investigators can focus on the tasks for which their expertise and experience are most needed such as interpretation of results, validation of hypotheses, etc. Moreover, SADFC implements mechanisms allowing to satisfy law requirements. These mechanisms rely on formal definitions on which this paper focuses: a formalization of the event reconstruction containing formal definitions of entities involved in an incident and four sets of operators allowing to extract, manipulate and analyse the knowledge contained in the model. The case study presented in the previous section has shown the relevance of this model. Indeed, the use of a semantically rich representation using new semantic aspects (in addition to temporal aspects) allows to build advanced analysis processes. In particular, we have presented the possibilities introduced by our model to correlate events. The identification of correlated events enables to highlight valuable information for the investigators.

Future works will be to develop the two other components of our approach. First, we will design an investigation process model providing a framework for the development of automated investigation tools. This process model will include a definition of the steps composing a digital investigation, from the preservation of the crime scene to the presentation of conclusions to Justice passing through the definition of the crime scene, the collection of footprints, the construction and the analysis of the timeline. During the development of this model, the focus is on the precision and completeness of the definition of each step as well as its inputs and outputs. Second, we will implement the theoretical model presented in this paper. This implementation will consist of a reference architecture based on an ontology derived from our knowledge model coupled with inference mechanisms and analysis algorithms. This architecture will be used to validate our model.

## Acknowledgements

The above work is a part of a collaborative research project between the CheckSem team (University of Burgundy and the UCD School of Computer Science and Informatics) and is supported by a grant from University College Dublin and the Burgundy region (France). The authors would like to thank Dr Florence Mendes for valuable comments on the formal part of this paper.